\documentclass[10pt,twocolumn]{article}
\usepackage[top=20mm,left=20mm,right=20mm,bottom=20mm]{geometry}

\usepackage[utf8]{inputenc}
\usepackage[T1]{fontenc}
\usepackage{graphicx}
\usepackage{longtable}
\usepackage{float}
\usepackage[normalem]{ulem}
\usepackage{hyperref}
\usepackage{wrapfig}
\usepackage{wrapfig}
\usepackage{soul}
\usepackage{textcomp}
\usepackage{marvosym}
\usepackage{wasysym}
\usepackage{latexsym}
\usepackage{amssymb}
\usepackage[table,usenames,dvipsnames]{xcolor}
\usepackage{cite}
\usepackage{verbatim}

\usepackage[font=small,labelfont=bf]{caption}

\usepackage{times}

\usepackage{subcaption}
\usepackage{url}

\usepackage{gensymb}

\usepackage{authblk}

\newcommand{\westermo}{Westermo}
\newcommand{\westermonct}{Westermo Network Technologies AB}

\title{Intermittently Failing Tests in the Embedded Systems Domain}

\author[1,2]{P. E. Strandberg}
\author[2]{T. J. Ostrand}
\author[2,3]{E. J. Weyuker}
\author[2]{W. Afzal}
\author[2]{D. Sundmark}
\affil[1]{\small{Westermo Network Technologies AB, Västerås, Sweden}}
\affil[2]{\small{Mälardalen University, Västerås, Sweden}}
\affil[3]{\small{University of Central Florida, Orlando, USA}}

\date{\small{Accepted to the ACM SIGSOFT International Symposium on Software Testing and Analysis (ISSTA), 2020.}}

\begin{document}

\maketitle

\section*{Abstract}
Software testing
is sometimes plagued with intermittently failing tests and finding the root causes of such failing tests is often difficult. This problem has been widely studied at the unit testing level for open source software,
but there has been far less investigation at the system test level, particularly the testing of industrial embedded systems.
This paper describes our investigation of the root causes of intermittently failing tests in the embedded systems domain, with the goal of better understanding, explaining and categorizing the underlying faults.
The subject of our investigation is a currently-running industrial embedded system, along with the system level testing that was performed. 
We devised and used a novel metric for classifying test cases as intermittent. From more than a half million test verdicts, we identified intermittently and consistently failing tests, and
identified their root causes using multiple sources. 
We found that about 1-3\% of all test cases were intermittently failing.
From analysis of the case study results and related work, we identified nine factors associated with test case intermittence.
We found that
a fix for a consistently failing test typically removed a larger number of
failures detected by other tests than a fix for an intermittent test.
We also found that more effort was usually needed to identify fixes for intermittent tests than for consistent tests.
An overlap between root causes leading to intermittent and consistent tests was identified.
Many root causes of intermittence are the same in industrial embedded systems and open source software.
However, when comparing unit testing to system level testing, especially for embedded systems, we observed that the test environment itself is often the cause of intermittence.

\phantom{Invisible vertical space}

\noindent%
\textbf{Keywords:}
system level test automation,
embedded systems, 
flaky tests,
intermittently failing tests,
non-deterministic tests

\section{Introduction}
The software in embedded systems has to be tested under realistic conditions, using real hardware~\cite{banerjee2016testing, wolf1994codesign}.
Even when using continuous practices, such as nightly testing, system testing
is a resource constrained process when compared to unit level testing,
and there is typically not enough time to execute all test cases every night on all hardware versions~\cite{maartensson2016continuous, strandberg2016}.
If test cases fail intermittently in nightly testing,
troubleshooting is often very costly and reproduction of 
failures very difficult, leading staff to
distrust test results~\cite{luo2014empirical}.

Intermittently failing hardware devices
have been studied for at least 70 years~\cite{cooper1947electrical}.
As software-based functionality has become increasingly important in large systems, non-deterministic results have continued to be a problem~\cite{ball1969effects, malaiya1979survey, leveson2004role}.
With improved automation and continuous integration, the volume
of test results exacerbates the problem~\cite{shahin2017continuous,strandberg2019flow}.
This problem has been studied in related work 
as described in Section 2,
focusing primarily
on unit testing level of open source software. 
The conclusion has generally been that test cases of poor quality fail unpredictably, even when testing the same software, and that
the test cases are to blame for intermittent failures~\cite{luo2014empirical, vandeursen2001refactoring}.
This paper continues
this research, but focuses on industrial \textit{embedded systems} being developed and tested at the \textit{system level}
with evolving software and testware.
In this type of environment we hypothesize 
that intermittently failing tests may have root causes 
that can stem from the testware, hardware, software, or their interfaces.

The \textbf{research objectives} of this paper are to
identify intermittently failing tests during system level testing by using an easily computed measure of the frequency of change of test verdict.
Additionally we want to find, explain and categorize the root causes of such failures in the context of 
development and maintenance of industrial embedded systems
designed by using continuous integration, and to compare our findings with similar research to build knowledge in the topic.

The \textbf{main contributions} of this paper are:
\begin{enumerate}
\item Nine
factors that may lead to intermittently failing tests for system level testing of an embedded system:
test case assumptions,
complexity of testing,
software or hardware faults,
test case dependencies,
resource leaks,
network issues,
random number issues,
test system issues, and
code maintenance.

\item Definition of a metric that can identify intermittently failing tests, and can measure the frequency of intermittence over the test base of a system.

\item Evidence that failures detected by intermittently failing tests often
require more work to find their root cause than those detected by consistently failing tests. This may lead testers to give up without identifying the failure's root cause, and to rely on the hope that failures will occur infrequently enough that they can be ignored.

\item Experience that a fix for a consistently failing test often repairs
a larger number of failures found by other test cases than a fix of an intermittently failing test.
\end{enumerate}

The paper is organized as follows.
Section~\ref{section-terminology} presents terminology.
Section~\ref{section-examples} describes two real cases of intermittent tests that motivate our research.
Section~\ref{sec-modeling-intermittent-tests} introduces a novel metric for measuring intermittence,
which is then used to analyze data collected in the case study presented in Section~\ref{section-case-study}.
Section~\ref{section-intermittent-results} presents the results of the case study, and
Section~\ref{section-synthesis-and-related-work}
shows how our results differ from, confirm, and extend previous work.
Sections~\ref{section-intermittent-discussion} and~\ref{section-intermittent-conclusions} discuss our findings and our conclusions.

\section{Intermittently Failing Tests}

\subsection{Terminology}

\label{section-terminology}

In addition to the obvious hardware (HW) and software (SW) that comprise an embedded system (ES), development and testing also involve \emph{testware} (TW), which itself has both software and hardware components.  
The TW software includes items such as test harnesses and stubs, servers, test libraries, test scripts, automated test cases and code to control the automation.
TW hardware is the physical environment on which test cases are executed, including servers, cables, and peripheral equipment such as load generators.

The hoped-for result of running a regression test suite is that all tests in the suite will be successful, thus confirming that recent modifications to the system have not broken any existing functionality or property.  
With \emph{continuous integration} (CI) methods, changes may occur at any time, and regression testing 
may be carried out daily, or even more frequently, to assure that recent changes have not adversely affected the system's behavior. 

During the development and maintenance of a large industrial embedded system, we have observed many examples of 
test cases that produce different verdicts in successive regression testing runs, with sequences of mixed verdicts.
Such results occur even for test cases that have no apparent connection to system modifications that were made since the previous test run. 

\textit{Flaky tests} are
tests that yield differing verdicts when nothing in the SW, HW or TW 
have been changed~\cite{luo2014empirical}.
The different verdicts occur because of hidden changes in the system state or the application's environment.
State changes can be caused by previously run test cases or by normal operation of the system.
Environment changes can occur spontaneously, and cause problems when system designers have failed to consider the possibility of their occurrence. 
Tests can also be flaky because of poor design.
These are sometimes called \textit{smelly tests}~\cite{fowler2018refactoring,garousi2018smells}.

In many industrial contexts, including the one that is the subject of this case study,
it is not useful to look for flaky tests, due to frequent changes made in the underlying SW, HW and TW in a CI development paradigm.
Since the perceived business value of retesting is limited when nothing has changed, we cannot expect to see much repeated testing on 
unchanged SW, TW and HW. 
Thus, although potentially flaky tests may exist, they are rarely observed and in fact represent a lack of understanding of all of the factors that might impact system behavior.
Bell et al.~\cite{bell2018deflaker} developed a tool to detect flaky tests without retesting that relies on instrumentation (code coverage), but this may be hard or impossible to use for CI, or for testing SW in resource constrained embedded systems~\cite{elbaum2014techniques}.

For these reasons, we define
an \textit{intermittently failing test} to be a test case that has been executed repeatedly while there is a potential evolution in SW, HW or TW, and where the verdict changes over time. 
Note that both flaky  and intermittently failing tests refer to the dynamic execution of test cases, and
 require that the system be executed at least twice before we can label them as such.
Furthermore, in automated system level testing of ES, intermittently failing tests are different from flaky tests in that they allow changes in the SW or HW of the ES under test, as well as in the TW used for testing.

A reader might object to our definition and argue that when changing an ES one should not be surprised that verdicts from test cases change as a result.
Although this is of course a relevant comment, in this paper we do not distinguish between 
expected and unexpected intermittently failing tests -- we are interested in these tests regardless of root cause.
A reader might also object to test cases being ``the same'' test cases, if the underlying test script code has been modified.
This is also a relevant comment, and as we will see in Section~\ref{section-synthesis-and-related-work}, code maintenance is indeed a factor that impacts intermittence, but not a dominating factor.

\begin{figure*}[th!]
  \begin{center}
  \begin{subfigure}[b]{0.18\linewidth}
    \centering
    \includegraphics[width=\linewidth]{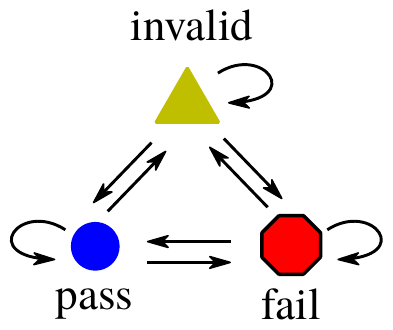}
    \caption{
      Markov chain
      \label{fig-q-score-chain}
    }
  \end{subfigure}
  \hfill
  \begin{subfigure}[b]{0.12\linewidth}
    \centering
    \includegraphics[width=\linewidth]{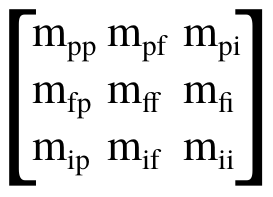}
    \caption{
      M
      \label{fig-q-score-m}
    }
  \end{subfigure}
  \hfill
  \begin{subfigure}[b]{0.33\linewidth}
    \centering
    \includegraphics[width=\linewidth]{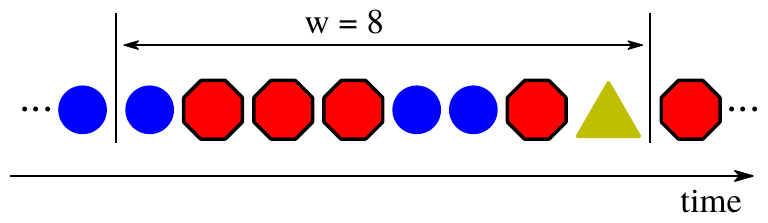}
    \caption{
      Observations
      \label{fig-q-score-observations}
    }
  \end{subfigure}
  \hfill
  \begin{subfigure}[b]{0.11\linewidth}
    \centering
    \includegraphics[width=\linewidth]{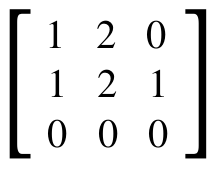}
    \caption{
      N
      \label{fig-q-score-n}
    }
  \end{subfigure}
  \hfill
  \begin{subfigure}[b]{0.18\linewidth}
    \centering
    \includegraphics[width=\linewidth]{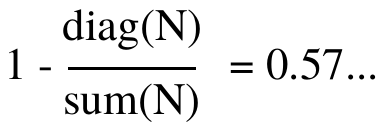}
    \caption{
      q-score
      \label{fig-q-score-equation}
    }
  \end{subfigure}
  \end{center}
  \caption{
    We model the test cases as Markov chains.
    The test case has some unknown transition probabilities, shown in matrix M.
    Counts of the 7 observed transitions in a window of length 8 are in matrix N, and
  the observed intermittence is computed with the q-score.
  Details described in Section~\ref{sec-modeling-intermittent-tests}.
      \label{fig-q-score}
      }
\end{figure*}

\label{subsection-q-score}

\subsection{Two Examples}
\label{section-examples}

\subsubsection{Example 1: A Smelly, Flaky and Intermittently Failing Test}
\label{example-room-temperature}
One test case at the company targets the temperature sensor that measures the internal temperature of the HW, and reports the value to other software.
If the SW receives a plausible value, 
the test case assumes that the  sensor is working properly.

The test was designed around the assumption that the ambient temperature would be between 20 and 40\degree C with an acceptable internal temperature between 21 and 60\degree C. This assumption was valid for several years and the test hardly ever failed. 
However, at one time the test lab was moved to a different location, and the test started to fail intermittently on one of the test systems. The new test lab had much better air conditioning, with one particular test system placed very close to one of the air conditioners. 
If HW in this test system had been powered off for some time, allowing it to cool down, and if the temperature sensor test was early in the test suite, prior to when the HW had heated up, then the test would fail because the internal temperature reported by the sensor was outside the plausible range.
Since these two conditions rarely occurred simultaneously, the failure was rarely observed. 
After a discussion with HW engineers, the plausible range was changed to 15 to 50\degree C, and the test stopped failing.

One could argue that this test case was smelly, because of the false underlying assumption of what an acceptable temperature range would be.
The test case was also intermittently failing.
However, in order to truly be able to label the test as flaky, we would have to investigate whether or not the test case had been executed at least twice with the same SW, HW and TW combination while also producing different verdicts. 

The temperature sensor test shows an almost trivial example of a fix, but it still illustrates the effort needed in identifying and dealing with intermittently failing tests: first, test results have to be observed, discussed and understood. Then an informed fix has to be applied -- a prerequisite for this particular fix was a discussion across roles (test engineer and HW engineer). Once the fix is applied, the test results have to be observed again for some time, to see that the issue has been properly addressed and there are no more failures.

This example also illustrates that ``the system state'' when testing an ES at system level 
can be very complex: the state contains not only all variables of the Linux kernel, the many physical components and possible flaws in soldering of the device, the test scripts and the order in which test cases are executed.
In this particular example, the system state also includes room temperature, and the internal temperature of the devices.
These parameters were not fully considered when the test script was first written.
Sometimes,
we cannot observe or control all variables in the system state, because we do not even know all of the factors that might impact the system's behavior.

\subsubsection{Example 2: Resource Leak in SW}
\label{example-intermittent-resource}
Another example further illustrates the complexity of automated testing of ES at the system level.
In one of the features in the SW under study, there was a bug in the form of a resource leak.
This feature was exercised by several test cases that could all trigger the leak.
But in order to observe the bug, a certain number of the test cases had to be executed for the resource to be depleted.
At the company, there is a regression test selection system in place that generates new test suites every night,
and one of the factors that leads to an increased priority of a test case is that it has  failed recently 
~\cite{strandberg2016,strandberg-nutshell}.
Time for testing is a scarce resource,
and testing terminates once time has run out, potentially leaving test cases that are late in the suite unexecuted. 
Therefore, if a certain test case passed recently, it might have a relatively low priority and not be executed at all during nightly testing.
But if the test case did fail, thereby identifying the bug, it would be prioritized and placed early in the suite. 
However, in the beginning of the suite it was unlikely that
the leak had been triggered enough for the test to fail.
In retrospect, we know that the leak was consistent, but because test suites changed over time, we did not consistently trigger the SW bug. We thus observed an intermittently failing test case where we could have observed a consistently failing test case if the suites had been identical over time.
This is a very common issue when there are resource leakages, and one of the reasons that these sorts of faults are so difficult to find and fix properly. Essentially, the leakage will only occur over a long period of time or when an unusual surge in workload has occurred.
If nightly testing begins in a standard or quiescent state, it might be that the failure is observed very rarely and only when a particular test case order occurs.

\subsection{A Model for Intermittently Failing Tests}
\label{sec-modeling-intermittent-tests}

We want to define a property of test cases that captures the
likelihood of a test failing intermittently, regardless of reason.
To model the sequence of varying test case results, we use the concept of a \emph{Markov chain} \cite{privault2013understanding}, which is a process with well-defined states, and probabilities for transitions from one state to another that depend only on the current state and input.
When run, test cases may result in one of three verdicts: pass, fail or invalid, where 
invalid means that the execution could not be completed.
This is typically triggered by unexpected behavior in the HW, SW or TW, followed by an unhandled exception in the TW.
We can think
of the test case verdicts (pass, fail, invalid) as being states in a Markov chain
(illustrated in Figure~\ref{fig-q-score-chain}), and each execution of the test case as a transition.
When a test verdict is different from the previous one, the transition is to a different state; a repeated verdict is represented by a transition to the same state.
Each test case has some unknown probabilities for
the transitions, as shown in a transition matrix $M$ (see Fig.~\ref{fig-q-score-m}).
The probabilities may change over time because of changes in the SW, TW, and HW,
and in practice we do not know what they are.

Based on historic data, we can observe the actual verdicts
that have been produced (Fig.~\ref{fig-q-score-observations}).
By counting transitions that occur in a sequence of consecutive test case executions, we construct the matrix $N$,
in which each element represents the number of observed
transitions from state to state
(Fig.~\ref{fig-q-score-n}).
While the elements $m_{i,j}$ of $M$
are the unknown \emph{probabilities} of going from $i$ to $j$, the
$n_{i,j}$ are the
\emph{number of observed transitions} from $i$ to $j$, and are dependent on the sequence of test cases actually run.  For example,
$n_{p,f}$ is the number of observed transitions from pass to
fail, corresponding to the number of times that a pass verdict is immediately followed by a fail verdict. The diagonal elements of $N$ represent how
often the test case verdict has remained the same, and the
off-diagonal elements represent how often the test case verdict  has changed, i.e., how often the Markov chain changes state.
We now define the \textit{q-score} as the
fraction of results in which the test case changes state, i.e.
$q = 1 - diag(N)/sum(N)$, (Fig.~\ref{fig-q-score-equation}).

The q-score is a direct measure of the variability of a test case's verdicts.
We have found it to be most useful when continuously measured over a moving window of approximately 8-15 consecutive verdicts of a test case.  A running plot of the score shows the varying intermittence over time of the test case, and shows whether the intermittence is increasing or decreasing.  Measurement over long periods such as several months can be used to identify test cases that have intermittency somewhere in their history, but is not useful for determining their most recent status.

Figure~\ref{fig-q-score} illustrates a window size of eight, giving seven
observed transitions.
The test
case remains in the same state three times (one pass $\rightarrow$ pass,
two fail $\rightarrow$ fail) and changes state four times, indicating that
the test case is more likely to change verdict than to retain
the verdict. The computed q-score for this window is 0.57.
Given a threshold value of the q-score, one could now use
this metric as a method for detecting intermittently failing tests.
We can also
compare the observed intermittence of test cases,
e.g.,\ a test with a q-score of 0.57 could be thought of as more intermittent
than a test case with a q-score of 0.15.
Thus rather than simply deciding that a test case is either intermittently failing or not, much as the literature on flaky tests has done, we can instead use the q-score as a way of deciding which tests are most problematic because of their changing behavior.
Of course, it may be that test cases exhibit a high q-score for totally expected reasons such as when the SW or test scripts 
are being refactored, or undergoing code maintenance
or when the HW is being replaced.
One could therefore consider combining the q-score with other metrics, such as code churn.

It is also 
useful
to introduce a metric for how frequently a test case passes.
We define the p-score as the fraction of passed verdicts in a window of observed verdicts.
Using the same example as above (Figure~\ref{fig-q-score}),  $p=3/8=0.375$,
indicating that this test case passed in 37.5\% of the executions.

Both the windowed p-score and q-score will be used to select test cases of interest from
a large pool of possible test cases in the coming sections.
We will also use the overall scores from all verdicts for each test case to
describe the intermittence of test cases on a population level.

As we will demonstrate in our case study, different q-score thresholds and window sizes will select different test cases as potentially problematic.
The q-score's range of values from 0 to 1 provides the system tester a more precise means of setting the degree of variability to indicate which test case results are most urgent to investigate, in contrast to a binary metric such as flaky tests.

\begin{figure*}[t!]
    \centering
    \includegraphics[width=0.95\linewidth]{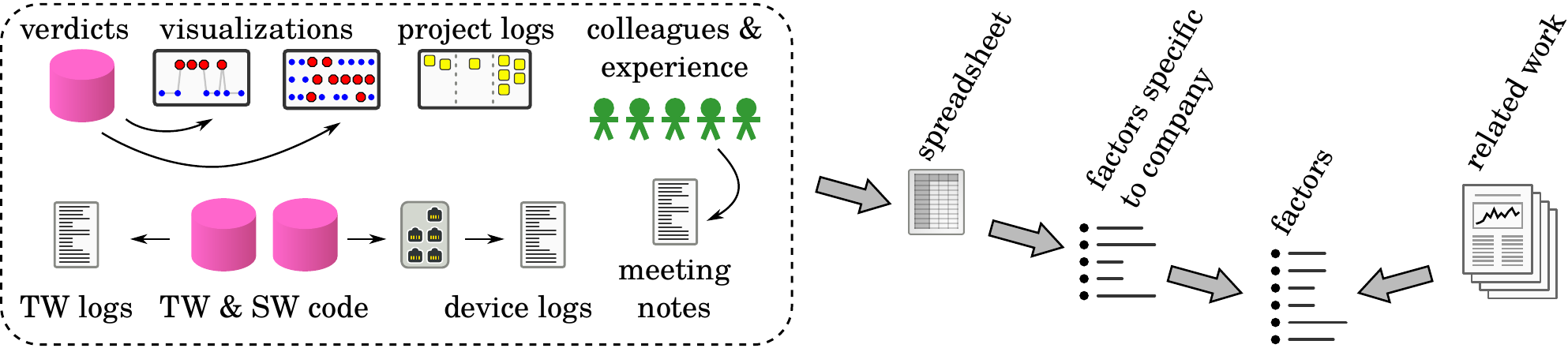}
    \caption{
      Using several sources of information, we collected information on root causes and fixes for intermittently failing tests.
      With related work we created a list of relevant factors.
      \label{fig-research-process}
    }
\end{figure*}

\section{Case Study Design}
\label{section-case-study}

This section describes the research questions, the case, and the industrial context, 
as well as data collection, and data analysis procedures.
The overall research flow is illustrated in Figure~\ref{fig-research-process}.

Our goal is to explore and explain the root causes of intermittently failing tests observed during system level testing in the development and maintenance of embedded systems developed using continuous integration in an industrial environment. We will compare these test cases with those that consistently cause  failures.
We formulate two research questions (RQs):

\begin{itemize}
 \item \textbf{RQ1:} What are the root causes of intermittently failing tests when testing embedded systems on a system level?
 \item \textbf{RQ2:} Are the root causes different for tests that fail consistently?
\end{itemize}

\subsection{Case and Subject Selection}

The \emph{case} in this study is nine months of test results from nightly testing of an operating system for embedded systems.
We analyze the case using four units of analysis:
two groups of intermittently failing tests, and
two groups of consistently failing tests.
This study could thus be defined as an embedded exploratory and explanatory case study with four units of analysis~\cite{runeson2012}.

\subsubsection{Industrial Context}
Recent research literature 
on flaky tests has typically focused on unit testing of open source projects targeting a general-purpose computer (not ES).
To broaden our understanding of intermittently failing tests in industrial systems, the case study
is conducted at
\westermonct\, 
an international company with about 250 employees targeting the on-board rail, track-side rail, power distribution,
and other industrial automation domains where communication networks in harsh environments are needed.
The company develops switches and routers (the HW) running an operating system (the SW), which is developed in-house in an agile development process based on Kanban where new features are developed in separate code branches.
The company has invested heavily in test automation. The testware (TW) includes
an internal infrastructure for nightly testing,
equipment such as test servers and devices under test with a combined weight of more than a ton,
and a test framework with hundreds of test scripts, to which new ones are added weekly.
Several person-years of effort have been invested in coding and construction of the TW.
SW development takes place in feature branches that are expected to receive many code changes, and
once a feature is implemented and stabilized, it is merged into a master branch.

To get an understanding of how frequently the same SW is tested on consecutive nights (if at all),
we analyzed how many nights in a row  the same SW, the same TW, and both the same SW and TW are used.
The median number of nights of regression testing with the same revision of the SW was 2.5 in the branch we investigated.
The corresponding number for TW was only 1 (see details in Table~\ref{table-same-same}).
In addition to code changes, there are many variables in the TW that are hard to observe, log, or control: network latency, room temperature where devices operate, states in test servers, etc.
When testing an embedded system at the system level, running only a subset of the regression test suite may be necessary, because testing takes far more time at the system level than testing done at the unit level.
Selecting a subset will typically lead to test cases being reordered in the test suites~\cite{strandberg2016,strandberg-nutshell},
which may trigger intermittent failures if there are test case dependencies.

\begin{table}[t]
  \caption{
     Number of consecutive nights of testing with the same SW, TW, or both.
    \label{table-same-same}
  }
  \begin{center}
    \begin{tabular}{@{}p{1in}ccccc@{}}
      Nights w.\ same &Min& Max & Avg. & Med.& Std.d. \\
      \hline
      SW Code         & 1 & 107 & 5.72 & 2.5 & 13.97 \\
      TW Code         & 1 &  28 & 3.02 & 1.0 &  3.96 \\
      SW and TW Code  & 1 &  17 & 2.32 & 1.0 &  2.68
  \end{tabular}
  \end{center}
\end{table}

The software quality process at the company includes
several types of testing techniques in addition to automated regression testing (such as manual risk-based testing, robustness testing, etc.),
as well as bug tracking, bug triage meetings, test results sync meetings,
and release gate meetings, before a new version of the SW is made available to customers.

\subsection{Data Collection Procedures}

\subsubsection{Raw Data}

After discussions with \westermo, we extracted verdicts from 270 consecutive nights of testing of a stable code branch.
This branch has existed for a long time, so we expected to see mostly passing tests. By using a stable code branch, the impact of changes in SW should be minimized, in contrast to a feature branch that is short-lived and has frequent SW changes. In such a case we would expect many failing tests.
The extracted data contains 
532069 verdicts from 5212 test cases.
In this study, we consider a test case to be a combination of one of the 13 test systems, 527 test scripts, and 69 different parameter settings in the raw data. Note that if  the same script is run on two different systems, we count it as two different test cases.

Each individual test case ran at most once each night, so had between 1 and 270 verdicts during this period.
One test case is hard coded into the beginning of the scripts,
so it ran every night. 
The average was 102.1 verdicts with a median of 96 and standard deviation of 40.3. This means that each night about a third of all test cases were included in nightly testing of this branch, with the particular test cases selected varying from night to night.

Of the 532069 verdicts, 526335 (98.9\%) were pass, 2673 (0.5\%) were fail, and 3061 (0.6\%) invalid.
For this study, we are not concerned with the impact of the failed and invalid verdicts, so there is no severity associated with the non-pass verdicts.

We removed the seven test cases that had only been executed once during the 270 night period.
Based on the complete sequence of verdicts for each remaining test case, we found that the tests have an average p-score of 0.986 and an average q-score of 0.011.
This means that the test cases could be thought of as having an overall probability of 1.1\% to change verdict in this data set.
The maximum q-score of 0.68 means that at least one or a few test cases were more likely to change verdict than to retain a verdict.
The median p-score of 1.0 and q-score of 0.0 indicates that almost all test cases never fail, and never change verdict.
24.8\% of the test cases had a non-zero q-score.
See details in Table~\ref{table-p-q-score}.

\begin{table}[t]
  \caption{Minimum, maximum, average, median and standard deviation of the q-scores and p-scores of all 5205 test cases that were executed more than once. \label{table-p-q-score}}
  \begin{center}
    \begin{tabular}{@{}lccccc@{}}
      Score       & Min &  Max &  Avg. & Med. &  Std.d. \\
      \hline
      p-score     & 0.0 & 1.00 & 0.986 & 1.0 & 0.060 \\
      q-score     & 0.0 & 0.68 & 0.011 & 0.0 & 0.037 \\
    \end{tabular}
  \end{center}
\end{table}

\subsubsection{Categorizing Intermittently and Consistently Failing Tests}
\label{four-groups}
 \begin{figure}[t]
   \centering
   \includegraphics[width=\linewidth]{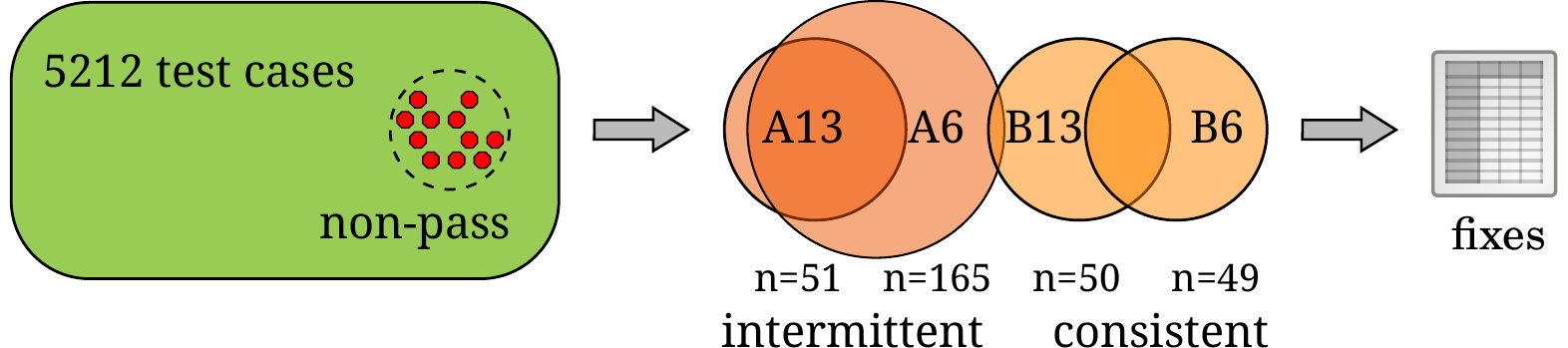}
   \caption{
     From all test cases we identified intermittent and consistent tests, and analyzed their fixes.
     \label{fig-test-cases-p-q-fix}
   }
 \end{figure}
Our goal is to identify and study fixes such that we can determine the root causes of intermittently and consistently failing tests.
From the 5212 tests in the database, we identified the following two groups of tests:

\begin{itemize}
\item Group A: test cases that at some point in time intermittently failed (i.e. had a high q-score), but after fixes were made, then mostly passed (i.e. had a high p-score).
\item Group B: test cases that at some point in time failed consistently (i.e. had a low p-score and also low q-score), but after changes were made, they mostly passed (i.e. had a high p-score).
\end{itemize}

To study changes in q-score and p-score over time we use moving windows of verdicts, as illustrated in Figure~\ref{fig-q-score}.
By using a small window we hope to capture rapid changes in q-score,
and with a larger window we hope to capture slower trends.
In general, we want the A and B groups to contain enough test cases to permit meaningful analysis, while not too many to be overwhelming. 
We generated many samples of the A and B groups, using window sizes ranging from 5 to 30 and different cut-off values of p and q. 
In the end, we used two window sizes for a total of four groups --
groups A6 and B6 used a window size of 6 verdicts, while A13 and B13 used 13 verdicts.
For all four groups we required that they ended up mostly passing, with a final p-score of at least of 0.96.
For group A13, we required a q-score of at least 0.35 at some point in time, and
for A6, we required a q-score of at least 0.5.
For B6 and B13, we required the p-score to go below 0.2 at some point in time.
With these window sizes and thresholds, we got a total of 230 test cases in the four groups.

To summarize, for inclusion in A, a test case must:
(i) at some point, have had a floating q-score over a certain limit
(0.5 for A6, and 0.35 for A13)
and
(ii) the final floating p-score must end above a certain limit
(0.96 for both A6 and A13).
These are thus test cases that have been intermittently failing, but end up passing
(examples are visualized in Figures~\ref{fig-example-a6} and~\ref{fig-example-a13}%
).
For inclusion in B, a test case must:
(i) not be in A%
\footnote{For inclusion in B, we applied the criteria of not being in A for a given window size, e.g.\ we allowed inclusion of a test case in both A6 and B13, but not in A6 and B6.
}, (ii) at some point have had a p-score below a certain limit
(0.2 for both B6 and B13),
and (iii) the final floating p-score must end above another p-score limit
(0.96 for both B6 and B13).
In other words, these are test cases that have been mostly failing for a period, but that end up passing
(an example is visualized in Fig.~\ref{fig-example-b13}%
).
This resulted in a total of 230 test cases,
with
165 test cases in A6,
51 in A13,
49 in B6 and
50 in B13.
Figure~\ref{fig-test-cases-p-q-fix} shows the large overlap between the test groups, as
49 of the 51 tests in group A13 were also present in A6,
34 of the 49 tests in B6 were also present in B13,
and two of the test cases were in both A6 and B13. No test case was in three or four groups.

\subsection{Analysis Procedure}
\label{intermittent-analysis-procedure}

\begin{figure*}[t]
  \begin{center}
  \begin{subfigure}[b]{0.31\linewidth}
    \includegraphics[width=\linewidth]{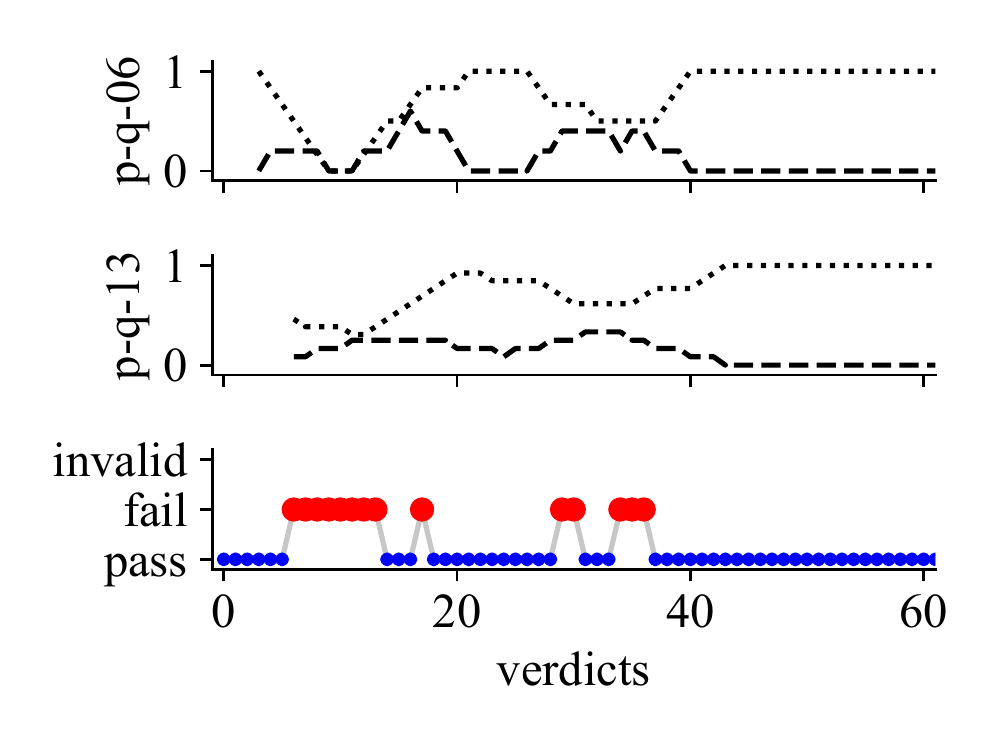}
    \caption{
      A6 
      \label{fig-example-a6}
    }
  \end{subfigure}
  ~
  \begin{subfigure}[b]{0.31\linewidth}
    \includegraphics[width=\linewidth]{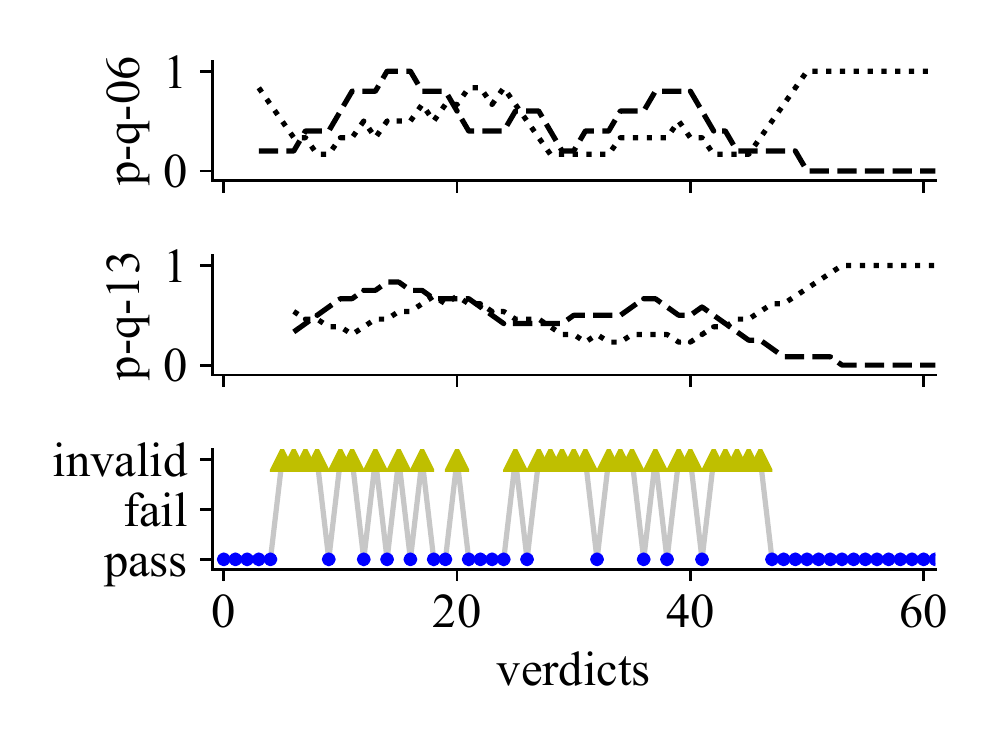}
    \caption{
      A13 
      \label{fig-example-a13}
    }
  \end{subfigure}
  ~
  \begin{subfigure}[b]{0.31\linewidth}
    \includegraphics[width=\linewidth]{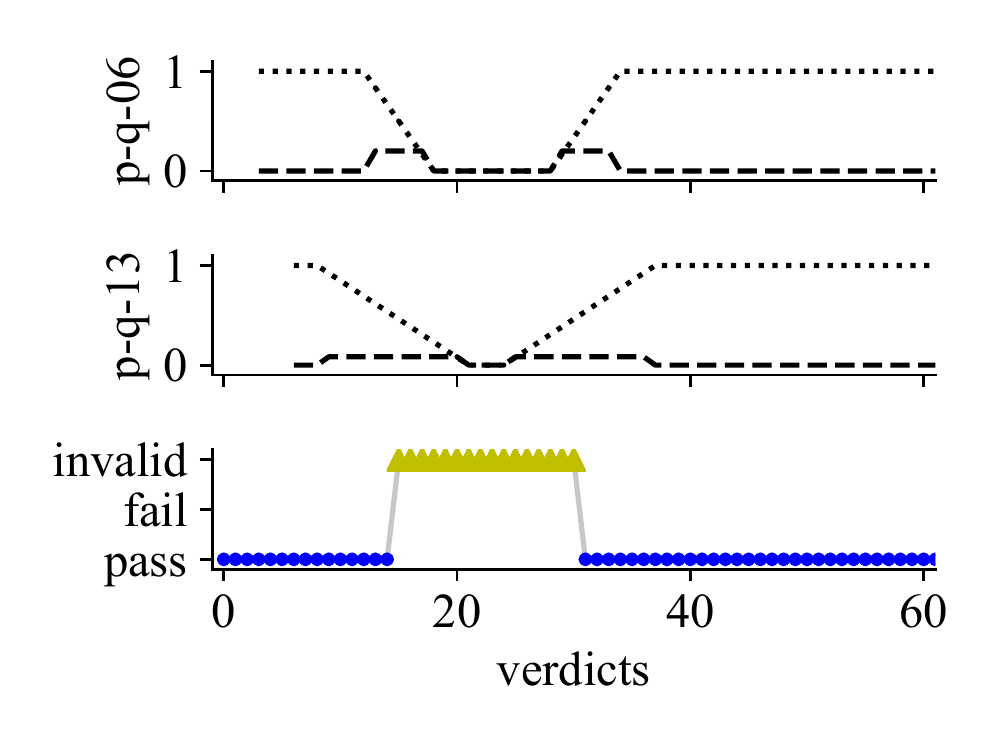}
    \caption{
      B13
      \label{fig-example-b13}
    }
  \end{subfigure}
  \caption{Verdicts over time from three test cases, as well as q-score (dashed) and p-score (dotted) for
  window size 6 and 13. 
    \label{fig-examples-a13-b6}}
  \end{center}
\end{figure*}

The 230 test cases of interest were analyzed.
Typically this investigation required more than one source of information and involved inspection of test and project artifacts,
or visualizations thereof.
The tools used were:
(i) A visualization of the verdicts from the test case, as illustrated in Figure~\ref{fig-examples-a13-b6}.
(ii) A second type of visualization, this one heatmap-like, that uses test results from before or after the date range we investigated. These plots
can quickly determine whether the test case is still under investigation, i.e.,\ if a test case was failing in a later phase and no
obvious fix has been found, then it was assumed that it was still under investigation.
Example visualizations of the heatmap plot used can be found in Figure~10 in~\cite{strandberg2018trdb}.
In addition to these visualizations, (iii) the SQL-interface to the test results database was also used for answering non-standard queries,
such as at what time a certain piece of HW was replaced in a test system.
Log files from (iv) the test framework, as well as test framework communication logs with (v) devices under test and (vi) peripheral equipment such as load generators in the test systems were manually inspected for error messages or other issues.
Project artifacts in (vii) issue trackers and (viii) planning tools
were also used,
as well as (ix) historic hand written notes from developer-tester sync meetings from the time range of investigation.
In some cases, the source code repository logs of (x) the test framework code or (xi) SW code were investigated.
In addition to these tools, scripts and artifacts, we also used (xii) personal experience of \westermo\ staff (including the first author).

For each of the 230 test cases, observations were hand-written on paper, resulting in 34 pages by the end of the study.
An anonymized summary was written in an on-line spreadsheet for simplified collaboration between authors.
For each test case we made notes on tools used, fix identified or root cause.
A typical example of the written notes is:
\begin{itemize}
  \item
   \textbf{Tools used:} test framework log file, heatmap, test framework source code changes.
  \item
    \textbf{Fix:} Revert of code change that applied the system configuration in incomplete steps,
    instead of a complete and correct configuration.
  \item
    \textbf{Root Cause: }Code change in test script that applies an intermediate (broken) configuration involving three features.
\end{itemize}

When all 230 test cases had been analyzed, we grouped the fixes or root causes into categories and subcategories (summarized in Table~\ref{table-categories-fixes} and discussed in Section~\ref{results}).

\section{Case Study Results} \label{results}

\label{section-intermittent-results}

\subsection{Fixes and Root Causes (RQ1)}

We identified five main categories of root causes:
(i) Changed HW allocation for testing,
(ii) test case assumptions,
(iii) test system issues,
(iv) SW or HW Faults,
and
(v) code maintenance of TW code.
In the analysis, we also identified
test cases still under investigation,
test cases that had more than one fix or root cause,
and test cases for which we were unable to identify a fix or root cause.
Table~\ref{table-categories-fixes} 
summarizes these categories.
In the following paragraphs, we discuss each category in detail.

\textbf{Hardware Resource Allocation:}
Seven fixes involved modifications of the HW allocation. This means testing with a different subset of the available physical devices, but with the same test script on the same test system (the HW selection algorithm is presented in~\cite{strandberg2018automated}).
Five of the fixes involved avoiding a link breaker that was used incorrectly (or not at all) by the test, but that could still interfere with traffic.
One test script required a port with nothing on the other end, and another required traffic to flow through the back plane of the device in ways not adhered to. In total, the seven fixes stabilized nine combinations of test script, parameter settings, and test system.

\textbf{Test Case Assumptions:}
Nine fixes repaired incorrect assumptions about the test framework and test systems, particularly assumptions involving timing (5 of 9).
In addition to changed intervals or tolerances on timing, one test script required a modified temperature range (as mentioned in Section~\ref{example-room-temperature}).
One fix involved a search for an event in the latest logged events in a log file in the HW, but due to other activities using the same log file, the test script had to expand the search to a larger portion of the file. 
Another test script assumed that exactly one instance of a unique type of HW was present in the test system, and could fail if more than one was present.
Yet another fix was to update the TW code that generates traffic.
The final fix in this category solved assumptions on the versions of a third party library used by the test framework.
In total, the 9 fixes stabilized 31 combinations of test script, parameter settings, and test system.

\begin{table}[t]
\caption{%
  Distinct categories of fixes.
  E.g., there was a total of 5 distinct test cases with HW allocation of linkbreaker as root cause, 4 of which were in A6, 4 were in A13 and 1 in B6.\label{table-categories-fixes}}
\begin{center}\begin{tabular}{|l|r|r|r|r|r|}
\hline
Root Cause/Fix                   &  A6 & A13 &  B6 & B13 & Tot. \\
\hline
HW Allocation                    &   4 &   4 &   3 &   0 &   7     \\
 \emph{ - link breaker}          &   4 &   4 &   1 &   0 &   5     \\
 \emph{ - switch core}           &   0 &   0 &   1 &   0 &   1     \\
 \emph{ - empty port}            &   0 &   0 &   1 &   0 &   1     \\
\hline
TC Assumptions                   &   7 &   3 &   2 &   1 &   9     \\
 \emph{ - timing}                &   5 &   2 &   0 &   0 &   5     \\
 \emph{ - test system layout}    &   1 &   1 &   0 &   0 &   1     \\
 \emph{ - temperature}           &   1 &   0 &   0 &   0 &   1     \\
 \emph{ - log file}              &   0 &   0 &   1 &   0 &   1     \\
 \emph{ - lib.\ version}         &   0 &   0 &   1 &   1 &   1     \\
\hline
Test System Issues               &   9 &   6 &   2 &   2 &  11     \\
 \emph{ - replace device}        &   4 &   3 &   0 &   0 &   4     \\
 \emph{ - console junk}          &   2 &   1 &   0 &   0 &   2     \\
 \emph{ - I/O relay}             &   2 &   2 &   0 &   0 &   2     \\
 \emph{ - USB sticks}            &   1 &   0 &   0 &   1 &   1     \\
 \emph{ - FTP server}            &   0 &   0 &   1 &   1 &   1     \\
 \emph{ - license}               &   0 &   0 &   1 &   0 &   1     \\
\hline
SW or HW Faults                  &  16 &   2 &   0 &   0 &  16     \\
 \emph{ - SW impact on HW}       &   7 &   1 &   0 &   0 &   7     \\
 \emph{ - SW timing}             &   9 &   1 &   0 &   0 &   9     \\
\hline
Code Maintenance                 &   9 &   0 &   2 &   3 &  12     \\
 \emph{ - unclear}               &   7 &   0 &   0 &   1 &   8     \\
 \emph{ - broken renaming}       &   1 &   0 &   1 &   1 &   2     \\
 \emph{ - traffic generator}     &   1 &   0 &   0 &   0 &   1     \\
 \emph{ - forgotten patch}       &   0 &   0 &   1 &   1 &   1     \\
\hline
Multiple Root Causes             &   8 &   2 &   0 &   1 &   8     \\
\hline
Under Investigation              &  14 &  34 &   0 &   0 &  35     \\
\hline
Unknown Fix                      &   8 &   5 &   0 &   0 &   8     \\
\hline
\end{tabular}\end{center}\end{table}

\textbf{Test System Issues:}
Eleven fixes were modifications to the test systems: four of these involved replacing devices.\footnote{When developing new HW products, \westermo\ runs prototypes in nightly testing to verify SW-HW integration. Over time, one test system might use several generations of prototypes, as well as released HW products.}
Two were fixed by rebooting a server when junk characters were sometimes written in the console used between the TW and the HW.
In the test systems at Westermo, relays are used to power on and off devices.
Two of the fixes in A6 and A13 were related to worn out relays that had to be replaced (newer test systems use solid state relays to avoid this problem).
One fix was to complete the configuration of an FTP server that had not been finalized on a new test system.
Another fix was to insert USB-sticks that were assumed to be present in the HW, but had been forgotten, lost or broken.
Finally, one fix was related to a communication protocol that requires a license, but after replacing HW in the test system, the license was no longer valid.
In total, the 11 fixes stabilized 18 combinations of test script, parameter settings, and test system.

\textbf{SW or HW Faults:}
16 root causes were related to faults in the SW or HW.
Nine had a root cause in timing issues in the SW leading to intermittent failures,
and
seven were related to intermittent failures due to root causes in SW making a port in the HW temporarily unusable and thereby blocking traffic.
None of the SW or HW faults led to consistent failures -- apparently, any such fault had already been taken care of in features branches,
i.e.,\ these faults never reached the stable code branch we collected data from.
With duplicates, these 16 root causes accounted for a total of 53 combinations of test script, parameter settings, and test systems.

\textbf{Code Maintenance:}
A total of 49 test script, parameter, and test system combinations were related to refactoring or maintenance of TW code, with a total of 12 distinct fixes.
Two fixes had a root cause in renamed variables where not all instances of their use had been identified, which would lead to attribute errors
at run time.
Another fix was related to a new feature being developed in a separate feature branch of the SW. Once the feature was completed, several test cases were modified to account for some changes in behavior of the SW. However, not all test scripts that used this feature had been identified and modified. The remaining root causes coincided with maintenance or refactoring of the TW code, e.g., if many code changes had been added to a test case that failed intermittently, it was assumed to be undergoing maintenance.

\textbf{Under Investigation:}
35 root causes (and a total of 53 combinations) were still under investigation,
almost all of which were related to A6.
Two identified root causes seemed to be related to
other intermittent problems in the TW,
and TW assumptions on HW performance leading to undesired reboots of HW.

\textbf{Unknown and Overlapping Root Causes:}
For eight test cases, we found no obvious fix (with 9 total combinations of test script, parameters, and test system).
Another eight had two or more overlapping root causes (total: 8 combinations).

\textbf{Duplicated Fixes:}
If two test cases were fixed by the same fix or seemed to have the same root cause, they were considered duplicates,
but if similar fixes had to be applied several times, we treated them as separate fixes.
More than half of the combinations of test script, parameter settings, and test systems were duplicates (124 of 230).
The most duplicated subcategories were
SW or HW faults, under investigation, 
code maintenance,
and test case assumptions; only 10 of the 124 duplicates belonged to other sub-categories.

\subsection{Differences Between Intermittent and Consistent Tests (RQ2)}

The most important
differences we observed between (A) the intermittently failing tests, and (B) the consistently failing tests,
were that the B-group contained no test case with unknown root causes, no test case with SW or HW faults, and no test cases that were still under investigation.
On the other hand,
the A-groups had several test cases in these categories.
This provided evidence that confirmed our intuition 
that intermittency often made fault diagnosis significantly more difficult.
Another difference is that the B groups have a larger number of duplicates:
group B6 had 9 non-duplicates and 40 duplicates, meaning that 9 fixes would resolve all 49 non-passing test cases in B6.
In contrast, group A13 had only about 30\% duplicates, meaning that a fix for an individual test case was not very likely to also fix another test script, parameter, and test system combination.

The different fixes and root causes required different amounts of effort to identify,
using the tools explained in Section~\ref{intermittent-analysis-procedure}.
On average, we used 1.1 tools during the investigation when finding a duplicate, and
2.9 tools when finding a SW or HW fault.
During the investigation, unknown fixes required 4.5 tools,
and the remaining six categories between 2.2 and 2.6 tools.
The average number of tools required was 2.0 for both A6 and A13, and 1.2 for both B6 and B13.
This supports the hypothesis that a greater effort is needed to find the root cause of an intermittently failing test.

\section{RQ1 Revisited in Light of Related Work}
\label{section-synthesis-and-related-work}

Despite starting with a data set including half a million verdicts, our findings obviously do not provide a complete picture of intermittently failing tests in the embedded systems domain, since we only investigated a single system at a single company.
In this section, we revisit RQ1 and analyze our findings with respect to previous research.
In particular, we have explored
literature on intermittence in embedded and electrical systems,
including a 70 year old paper,
literature on how embedded systems are tested,
as well as recent literature on flaky  tests.

We have  found only four papers in which intermittently failing tests have been investigated at the system level:
Ahmad et al.~\cite{ahmad2019empirical},
Eck et al.~\cite{eck2019understanding},
Lam et al.~\cite{lam2019root}, and
Thorve et al.~\cite{thorve2018empirical}.
These papers report on intermittently failing tests in industrial embedded systems, Mozilla programs in various operating systems, Microsoft programs, and Android apps.
In particular, they report that intermittently failing tests are not uncommon, and that they represent a real problem. 
Based on our findings in the current case study, these four papers and other related work, we made some general observations,
and identified nine factors relevant for intermittently failing tests in the embedded systems domain.

\textbf{General Observations:} 
The term \emph{flaky tests} for the domain of regression testing and continuous integration was popularized in blog posts, e.g.~\cite{fowler2011eradicating}.
Luo et al.~\cite{luo2014empirical} did an investigation on unit tests in open source projects.
They found that many tests were flaky because of asynchronous waiting, concurrency, or test order.
Important problems with flaky tests are  they can be hard to reproduce, may waste time/involve maintenance cost, they can hide other bugs and they can reduce the confidence in testing such that practitioners ignore failing tests~\cite{luo2014empirical,vahabzadeh2015empirical}.
The findings of our case study support the generalization of  many of these findings to the system level of embedded systems.
We saw that finding the root cause for intermittently failing tests is frequently more difficult than for consistently failing tests, and
that one fix for a consistently failing test typically repairs several other issues,  whereas a fix of an intermittently failing test repairs fewer others.
We also saw that some test cases could have overlapping root causes, i.e., they were intermittently failing for more than one reason,
which supports the idea that intermittently failing tests may hide bugs.

\textbf{Measuring Intermittence:}
The q-score metric is similar to the model presented by Breuer in 1973~\cite{breuer1973testing},
but discovered independently, and created for another domain
(regression testing of SW-intense embedded systems as opposed to describing faults in HW).
As far as we can tell, Breuer was first to use Markov chains to describe faults in a system (as opposed to modeling a system under test).
Another way of quantifying flakiness, based on entropy, was presented by Gao~\cite{gao2017quantifying}. This method requires code instrumentation (code coverage), and was evaluated using Java programs with 9 to 90 thousand lines of code.
One perceived advantage of Gao's metric is that it can be used to ``weed out flaky failures.''
Labuschagne et al.\ \cite{labuschagne2017measuring} investigated build failures in continuous integration environments of open source software. As part of their data analysis, they used a Markov chain similar to ours. They investigated open source software when a ``build'' (a test suite) fails, and their use of transitions between build states was used to identify when exactly one non-pass build had occurred.

\textbf{Factor 1: Test Case Assumptions.} 
Test case assumptions include issues such as poor ranges, tolerances, timing, or concurrency. This was noted by several papers on system level testing (e.g.,~\cite{eck2019understanding, lam2019root, thorve2018empirical}) and was also seen in our case study. 
Abbaspour Asadollah et al.~\cite{asadollah2017concurrency} investigated concurrency bugs in the domain of open source SW and found that about 4\% of bugs were related to concurrency, and that they were slightly more severe and required slightly more time to fix than other bugs.
Musuvathi et al.\ \cite{musuvathi2008finding} showed that 
some issues in concurrent programs could consistently be reproduced 
if threading was monitored and controlled.

\textbf{Factor 2: Complexity of Testing.} 
As illustrated in the example in Section~\ref{example-intermittent-resource}, testing an embedded systems can be a complex process,
in particular when compared to running unit tests without target HW.
Testing of embedded systems may require
HW testing, extra-functional properties testing, network testing, system testing,
test execution on one or more systems under test, systematic and exploratory test case design, etc.~\cite{asadollah2015survey, garousi2017embedded}.
Eldh et al.~\cite{eldh2007component} analyzed faults in a telecom SW and found that many faults are not discovered in unit level testing because of complexity in testing that developers do not always understand, and that
many of the reported faults were not related to SW.
Furthermore,
Ostrand and Weyuker \cite{ostrand1984collecting} found that different types of faults were uncovered during unit testing than were found during system testing of an industrial software system, concluding that both types of testing were essential.

Regardless of whether or not the complexity of the TW is a \emph{root cause} for intermittently failing tests, it can be an exacerbating factor.
It was noted as such by ~\cite{eck2019understanding, ahmad2019empirical, thorve2018empirical}, as well as in our case study under the HW allocation and test system issues categories.
Complexity is a challenge for testing an embedded systems in general, not only for intermittently failing tests~\cite{strandberg2019flow}.
To mitigate the complexity, Lam et al.~\cite{lam2019root} indicate that the use of log file analysis may be helpful,
Ahmad et al.~\cite{ahmad2019empirical} indicate that improved test results reporting may be helpful,
and our previous work draws similar conclusions for testing of embedded systems in general~\cite{strandberg2018trdb}.
Similarly,
Jiang et al.\ \cite{jiang2017causes} used text mining and comparisons of industrial test execution logs from system and integration level testing at Huawei-Tech Inc.\ to reduce the burden of determining the root cause of failing tests. The method was successfully deployed and used in industry.
Furthermore, Herzig and Nagappan \cite{herzig2015empirically} analyzed patterns in test cases at Microsoft to predict if a failed test case had a root cause in TW, with the goal of reducing development effort.

Mårtensson et al.~\cite{maartensson2016continuous} investigated problems and experiences when striving for continuous integration of embedded systems. They found that the test environment is a limited resource, leading to a tendency to construct many test systems with custom HW.
This in turn
may lead to tests being intermittent because of the increased maintenance burden.
They also observe that test cases that pass on simulated HW in a local build, are not guaranteed to pass when tested on real HW.
Wiklund et al.~\cite{wiklund2017impediments} came to similar conclusions for test automation in general.

\textbf{Factor 3: SW or HW Faults.} 
The challenge of intermittent faults in electrical systems has been known for a long time. The earliest study we have been able to find on the topic is from 1947 in which Cooper \cite{cooper1947electrical} investigated electrical control of dangerous machines.
In the 1960s, Ball and Hardie~\cite{ball1969effects} published a
paper describing intermittent failures. The authors write that their ``experience has shown that field failures \ldots\ tend to be intermittent in nature.''
A few years later, Breuer~\cite{breuer1973testing} investigated testing of intermittent faults in digital circuits.

A forty year old paper by Malaiya and Su~\cite{malaiya1979survey} investigated intermittent faults in integrated circuits. A similar study was made more recently by Bakhshi et al.~\cite{bakhshi2014intermittent}. Both Malaiya and Su, and Bakhshi et al.\ came to similar conclusions regarding intermittent faults caused by HW: temperature, loose connectors, bad soldering, corrosion, and voltage fluctuations. Bakhshi et al.\ also mentions possible root causes caused by SW, such as timing failures, processor loads, memory leaks, and disk error.

Some bugs, sometimes referred to as Mandelbugs~\cite{grottke2005classification}, cannot be observed without state build-up. This causes a possible delay between activation and symptom, may require non-trivial timing or a particular sequence for activation, or have a non-trivial error-propagation.
The root causes identified in the `SW or HW faults' factor of our case study could be described as Mandelbugs.
According to Cavezza et al.~\cite{cavezza2014reproducibility}, a large portion of faults in both mission-critical SW and open source SW are due to Mandelbugs.
Di Martino et al.\ ~\cite{di2014lessons} investigated repair logs from a super computer and found that it was not uncommon for SW faults to propagate from one node to another, whereas this was rare for HW faults.
They also found that
HW faults were more numerous than SW faults, but that HW faults were more rapidly repaired than SW faults.
Sycofyllos found that most SW-related fatal failures, regardless of domain, have a root cause involving not only SW, but \emph{both} SW and a user, or \emph{both} SW and HW~\cite{sycofyllos2016empirical}.

\textbf{Factor 4: Test Case Dependencies.} 
Test dependencies may lead to intermittently failing tests of embedded systems as well as for unit level testing.
This was also 
observed by e.g.,~\cite{eck2019understanding, ahmad2019empirical, thorve2018empirical, vahabzadeh2015empirical} which they called a test smell.
As was the case with Mandelbugs, this factor can be non-trivial for developers to identify, and several approaches
to address the dependencies have been proposed, e.g.\ in~\cite{gyori2015reliable, zhang2014empirically}.
Anecdotally, test case dependencies have been seen at Westermo, but were not observed in our case study.

\textbf{Factor 5: Resource Leaks.} 
Resource leaks may lead to intermittently failing tests of embedded systems, and
have been identified as a factor for leads to flakiness in unit level testing~\cite{ahmad2019empirical, eck2019understanding, luo2014empirical, thorve2018empirical, vahabzadeh2015empirical}.
Avritzer and Weyuker~\cite{avritzer1995automatic} found similar results for system level testing in the telecom domain.
This factor has also  been observed at Westermo, as mentioned in the motivational example in Section~\ref{example-intermittent-resource}, but was not seen in the case study.

\textbf{Factor 6: Network Issues.} 
Network issues have been identified as a factor for intermittently failing tests by~\cite{ahmad2019empirical, thorve2018empirical}.
Several of the fixes observed in the case study were related to incorrect use of link breakers in the test systems, which sometimes had a negative impact on network performance leading to intermittently failing tests.
This factor is related to both test case assumptions and the complexity of testing.

\textbf{Factor 7: Random Number Issues.} 
Incorrect use of random numbers leading to intermittently failing tests at the system level was identified by both~\cite{ahmad2019empirical} and~\cite{eck2019understanding}. Some test cases at Westermo use random numbers, but it was not identified as a factor for failing tests in the current case study.

\textbf{Factor 8: Test System Issues.} 
Test system issues are related to the factor of complexity of testing.
In our data, we saw that replacement of HW prototypes,
interference in or misconfiguration of the console communication,
or issues with I/O for powering HW could lead to intermittently failing tests.
Alégroth and Gonzalez-Huerta, and Vahabzadeh et al. investigate technical debt and bugs in TW~\cite{alegroth2017towards, vahabzadeh2015empirical}; they found that bugs and technical debt can be as present in TW as in other SW.
For system level testing, technical debt and bugs in TW could very well have impacts on a test system.
Wiklund et al.~\cite{wiklund2017impediments} made a literature study on impediments for software test automation and observed several challenges with respect to test systems in general, including
environment configuration and quality issues (including false positives, false negatives and fragile test scripts).

\textbf{Factor 9: Code Maintenance.} 
In 2001, van Deursen et al.~\cite{vandeursen2001refactoring} 
spoke of problematic tests
in a paper on refactoring TW.
As expected, we observed that refactoring and maintenance of test code could lead to intermittently failing tests as well as to consistently failing tests.
\section{Discussion}

\label{section-intermittent-discussion}

From our case study and related work,
we identified nine factors that could lead to intermittently failing tests when doing system testing of embedded systems, as well as a number of subcategories, e.g.\ hardware allocation and test system issues.
Many of the subcategories have been identified in previous work. 
The three top reasons for unit level flaky tests identified by Luo et al.~\cite{luo2014empirical}
were asynchronous wait, concurrency and test order dependency.
These overlap with the factors we identified for intermittently failing system tests. It is thus tempting to generalize findings for intermittently failing tests at the unit level to system level testing.
However, we also observed important differences between their study and ours:
they observed that most test cases were flaky when first written, which was not the case for us, and
most of the flaky tests they investigated were flaky independent of platform whereas we saw no such indication.
Instead, we observed that test cases that failed consistently seemed to do so on more than one test system.

We observed differences in root causes for test cases that fail intermittently and consistently.
In particular, most tests for which the root cause was 
code maintenance failed consistently.
In addition, test cases that failed consistently were more likely 
to have a shared root cause -- meaning that one fix would repair
several combinations of test script, parameter settings, and test system, than those that failed intermittently.
We also observed that test cases that failed intermittently required more effort in terms of number of fixes than consistently failing tests.
They also required more effort in terms of tools used in order to identify the root cause.
However, the differences in root causes for test cases that fail intermittently and consistently are not easy to reason about.
As an example, we identified poor \emph{timing} as a factor for intermittence.
It is far from trivial to argue that this, or any other root cause, would lead to either intermittent or consistent failures.
Could one conclude that \emph{all} test cases with \emph{any} range would lead to intermittence, or only test cases with \emph{poor} ranges, and if so, how does one identify such a range?
When a 
test case developer designs
a test case,
can they easily identify a good range from a poor range?
In our case study, ranges in timing and temperature led to intermittent failures, while others involving
the number of lines of log file to parse led to consistent failures.
Was this merely coincidental in our study, or is it because of some characteristics of those ranges?
We observe that root causes leading to failing tests are not easily identified as risk factors for intermittent or consistently failing tests -- only as risk factors for failing tests. We hope to explore in the future characteristics that help to identify these sorts of differences.

Lam et al.~\cite{lam2019root} investigated flaky tests in five projects and found that 0.8 to 8.4\% of the test cases were flaky,
whereas we found 0.98\% to 3.17\% of the test cases to be intermittently failing.
However, q-score as a method to identify intermittently failing has a far from perfect signal to noise ratio, e.g.,\ many of the identified test cases were still under investigation after the date range during which we collected data.
This might indicate that the thresholds of p-score and q-score should be refined, or that q-score should be used with other metrics, e.g.,\ code churn, or for other purposes, perhaps as an indicator of the level of intermittence of a test case in the last month of testing.
The number of fixed intermittently failing tests and the quality of this classification can be expected to vary with thresholds and window sizes used for p-score and q-score. Future work could investigate suitable levels further.

An interesting question of responsibility that Ahmad et al.~\cite{ahmad2019empirical} raise is whether or not a test case should include retries and thereby potentially ``cover up'' the flakiness, or instead expose an issue. Implementing a retry seems to be a common correction strategy~\cite{ahmad2019empirical,bell2018deflaker,thorve2018empirical}.
We believe that retrying is a risky and unwise strategy that will lead to field failures because it simply masks the problem. In effect it is ignoring a known issue, even if it occurs only sporadically.
Gao~\cite{gao2017quantifying} strives for filtering away results from intermittently failing tests, so that a developer can focus on the consistently failing tests.
Again, we argue that tests that retry, or tools filtering away flakiness, would hide ``real issues'' in SW or HW as opposed to ``irrelevant'' issues with TW.
We therefore believe that intermittently failing tests should be considered crucial clues for investigating bugs that appear only sporadically.

Our findings imply that practitioners encountering an intermittently failing test when developing an embedded system could ask themselves:
\begin{enumerate}
\item Is the test case making incorrect assumptions on ranges or timing?
\item Is the actual testing context well understood?
\item Do test cases use shared resources without proper cleanup?
\item Is there a resource leak in the SW?
\item Are there assumptions on network performance, availability, etc., that are not always met?
\item Are random numbers improperly used?
\item Are the subsystems of the TW (test framework, the HW, or other peripheral systems used) well understood and properly configured?
\item Is there ongoing code maintenance or refactoring?,
or
\item Are you observing an intermittent issue in SW or HW that needs investigation?
\end{enumerate}

\subsection{Validity Analysis}
\label{evaluation-of-validity}

This case study relies on several constructs that originate in non-academic literature, that have not been carefully  defined, or that use several overlapping definitions.
Central constructs are \emph{non-deterministic tests}, \emph{flaky tests} and \emph{intermittently failing tests}.
We also introduced the metrics \emph{q-score} and \emph{p-score}.
In particular, Ahmad et al.\ criticizes the term \emph{flaky tests} because the shortcomings that lead to failures are not always in the tests,
making the term misleading~\cite{ahmad2019empirical}.
Furthermore, different studies sometimes define a flaky test with one definition,
but collect data using a different metric (test cases that produce different verdicts the same week, or over 10 executions, etc.).
There is thus an important threat to \emph{construct validity}, not only in this study, but in much of the prior work on flaky and non-deterministic tests.
We defined q-score as a way to measure intermittently failing tests, and collected test cases of interest using it. This has the advantage of making the study more reproducible, but introduces the potential risk of using a metric that is not yet ``proven in use.''
The metric may also have improved the \emph{reliability} of the study since other researchers with the same test results data could collect test cases of interest in the same way.

The fixes were identified by the first author using a number of tools available at \westermo, and tools developed for this study.
Being one researcher in this process may be a risk to \emph{internal validity} because identified fixes could have been different if other researchers had done the same analysis with the same data. The authors have been working at or with \westermo\ and company data for years, so there has been a \emph{prolonged involvement} which would reduce the risk of poor internal validity.
By using several tools in the analysis, and looking at the same data from different perspectives,
there was a form of \emph{triangulation} of the phenomena which could have reduced the threat to internal validity.

Research results that are \emph{generalizable} can be used in other contexts, and typically industry case studies claim poor generalizability.
One should thus see this study as one part of the increasing body of knowledge on intermittently failing tests in embedded systems, and not as the complete picture.

\section{Conclusion}
\label{section-intermittent-conclusions}

We studied intermittently failing tests in system level testing of embedded systems in a continuous integration development model.
Using a novel metric, we identified groups of tests that failed intermittently and consistently.
We analyzed the root causes and fixes of these tests and identified nine risk factors for intermittently failing tests:
test case assumptions,
complexity of testing,
software or hardware faults,
test case dependencies,
resource leaks,
network issues,
random numbers issues,
test system issues, 
and code maintenance.
The most important differences between consistently failing tests and intermittently failing tests are that
the intermittent tests did not always have a root cause that could be identified,
intermittent tests were sometimes indicators of software or hardware faults,
some intermittent tests were still under investigation after the data range we collected data from,
and fixes for a consistently failing test would also often repair other tests.
We also observe that many
root causes of intermittence in system level testing are the same as for unit level testing.

\section*{Acknowledgments}
This work was sponsored by \westermonct,
 the Knowledge Foundation (grants
 20150277, 
 20160139, 
 and
 20130258); 
 the Swedish Research Council (621-2014-4925); 
 the Swedish Innovation Agency (MegaM@Rt2);
 Electronic Component Systems for European Leadership 
 (737494);
and
EU Horizon 2020 (grant 871319).

\begin{small}
\bibliographystyle{abbrv}
\bibliography{refs.bib}

\begin{thebibliography}{10}

\bibitem{asadollah2015survey}
S.~Abbaspour~Asadollah, R.~Inam, and H.~Hansson.
\newblock A survey on testing for cyber physical system.
\newblock In {\em IFIP International Conference on Testing Software and
  Systems}. Springer, 2015.

\bibitem{asadollah2017concurrency}
S.~Abbaspour~Asadollah, D.~Sundmark, S.~Eldh, and H.~Hansson.
\newblock Concurrency bugs in open source software: a case study.
\newblock {\em Journal of Internet Services and Applications}, 8(1):4, 2017.

\bibitem{ahmad2019empirical}
A.~Ahmad, O.~Leifler, and K.~Sandahl.
\newblock Empirical analysis of factors and their effect on test
  flakiness-practitioners' perceptions.
\newblock {\em Preprint arXiv:1906.00673}, 2019.

\bibitem{alegroth2017towards}
E.~Al{\'e}groth and J.~Gonzalez-Huerta.
\newblock Towards a mapping of software technical debt onto testware.
\newblock In {\em Euromicro Conference on Software Engineering and Advanced
  Applications}. IEEE, 2017.

\bibitem{avritzer1995automatic}
A.~Avritzer and E.~J. Weyuker.
\newblock The automatic generation of load test suites and the assessment of
  the resulting software.
\newblock {\em IEEE Transactions on Software Engineering}, 21(9), 1995.

\bibitem{bakhshi2014intermittent}
R.~Bakhshi, S.~Kunche, and M.~Pecht.
\newblock Intermittent failures in hardware and software.
\newblock {\em Journal of Electronic Packaging}, 136(1):011014, 2014.

\bibitem{ball1969effects}
M.~Ball and F.~Hardie.
\newblock Effects and detection of intermittent failures in digital systems.
\newblock In {\em Proceedings of the November 18-20, 1969, fall joint computer
  conference (AFIPS'69)}. ACM, 1969.

\bibitem{banerjee2016testing}
A.~Banerjee, S.~Chattopadhyay, and A.~Roychoudhury.
\newblock On testing embedded software.
\newblock {\em Advances in Computers}, 101:121--153, 2016.

\bibitem{bell2018deflaker}
J.~Bell, O.~Legunsen, M.~Hilton, L.~Eloussi, T.~Yung, and D.~Marinov.
\newblock De{F}laker: automatically detecting flaky tests.
\newblock In {\em International Conference on Software Engineering}. ACM, 2018.

\bibitem{breuer1973testing}
M.~A. Breuer.
\newblock Testing for intermittent faults in digital circuits.
\newblock {\em IEEE Transactions on Computers}, 100(3):241--246, 1973.

\bibitem{cavezza2014reproducibility}
D.~G. Cavezza, R.~Pietrantuono, J.~Alonso, S.~Russo, and K.~S. Trivedi.
\newblock Reproducibility of environment-dependent software failures: An
  experience report.
\newblock In {\em International Symposium on Software Reliability Engineering}.
  IEEE, 2014.

\bibitem{cooper1947electrical}
W.~F. Cooper.
\newblock Electrical control of dangerous machinery and processes.
\newblock {\em Journal of the Institution of Electrical Engineers-Part II:
  Power Engineering}, 94(39):216--232, 1947.

\bibitem{di2014lessons}
C.~di~Martino, Z.~Kalbarczyk, R.~K. Iyer, F.~Baccanico, J.~Fullop, and
  W.~Kramer.
\newblock Lessons learned from the analysis of system failures at petascale:
  The case of blue waters.
\newblock In {\em International Conference on Dependable Systems and Networks}.
  IEEE/IFIP, 2014.

\bibitem{eck2019understanding}
M.~Eck, F.~Palomba, M.~Castelluccio, and A.~Bacchelli.
\newblock Understanding flaky tests: The developer's perspective.
\newblock In {\em Joint Meeting on European Software Engineering Conference and
  Symposium on the Foundations of Software Engineering}. ACM, 2019.

\bibitem{elbaum2014techniques}
S.~Elbaum, G.~Rothermel, and J.~Penix.
\newblock Techniques for improving regression testing in continuous integration
  development environments.
\newblock In {\em International Symposium on Foundations of Software
  Engineering}. ACM, 2014.

\bibitem{eldh2007component}
S.~Eldh, S.~Punnekkat, H.~Hansson, and P.~J{\"o}nsson.
\newblock Component testing is not enough-a study of software faults in telecom
  middleware.
\newblock In {\em Testing of Software and Communicating Systems}, pages 74--89.
  Springer, 2007.

\bibitem{fowler2011eradicating}
M.~Fowler.
\newblock Eradicating non-determinism in tests (blog post).
\newblock \url{https://www.martinfowler.com/articles/nonDeterminism.html},
  2011.
\newblock Online, Accessed 2019-06-26.

\bibitem{fowler2018refactoring}
M.~Fowler.
\newblock {\em Refactoring: improving the design of existing code}.
\newblock Addison-Wesley Professional, 2018.

\bibitem{gao2017quantifying}
Z.~Gao.
\newblock {\em Quantifying Flakiness and Minimizing its Effects on Software
  Testing}.
\newblock PhD thesis, University of Maryland, 2017.

\bibitem{garousi2017embedded}
V.~Garousi, M.~Felderer, {\c{C}}.~M. Karap{\i}{\c{c}}ak, and U.~Y{\i}lmaz.
\newblock What we know about testing embedded software.
\newblock {\em IEEE Software}, 35(4):62--69, 2018.

\bibitem{garousi2018smells}
V.~Garousi and B.~K{\"u}{\c{c}}{\"u}k.
\newblock Smells in software test code: A survey of knowledge in industry and
  academia.
\newblock {\em Journal of systems and software}, 138:52--81, 2018.

\bibitem{grottke2005classification}
M.~Grottke and K.~S. Trivedi.
\newblock A classification of software faults.
\newblock {\em Journal of Reliability Engineering Association of Japan},
  27(7):425--438, 2005.

\bibitem{gyori2015reliable}
A.~Gyori, A.~Shi, F.~Hariri, and D.~Marinov.
\newblock Reliable testing: detecting state-polluting tests to prevent test
  dependency.
\newblock In {\em International Symposium on Software Testing and Analysis}.
  ACM, 2015.

\bibitem{herzig2015empirically}
K.~Herzig and N.~Nagappan.
\newblock Empirically detecting false test alarms using association rules.
\newblock In {\em International Conference on Software Engineering}, volume~2.
  IEEE, 2015.

\bibitem{jiang2017causes}
H.~Jiang, X.~Li, Z.~Yang, and J.~Xuan.
\newblock What causes my test alarm? automatic cause analysis for test alarms
  in system and integration testing.
\newblock In {\em International Conference on Software Engineering}. IEEE,
  2017.

\bibitem{labuschagne2017measuring}
A.~Labuschagne, L.~Inozemtseva, and R.~Holmes.
\newblock Measuring the cost of regression testing in practice: a study of java
  projects using continuous integration.
\newblock In {\em Joint Meeting on Foundations of Software Engineering}. ACM,
  2017.

\bibitem{lam2019root}
W.~Lam, P.~Godefroid, S.~Nath, A.~Santhiar, and S.~Thummalapenta.
\newblock Root causing flaky tests in a large-scale industrial setting.
\newblock In {\em International Symposium on Software Testing and Analysis}.
  ACM, 2019.

\bibitem{leveson2004role}
N.~G. Leveson.
\newblock Role of software in spacecraft accidents.
\newblock {\em Journal of spacecraft and Rockets}, 41(4):564--575, 2004.

\bibitem{luo2014empirical}
Q.~Luo, F.~Hariri, L.~Eloussi, and D.~Marinov.
\newblock An empirical analysis of flaky tests.
\newblock In {\em International Symposium on Foundations of Software
  Engineering}. ACM, 2014.

\bibitem{malaiya1979survey}
Y.~K. Malaiya and S.~Y. Su.
\newblock A survey of methods for intermittent fault analysis.
\newblock In {\em International Workshop on Managing Requirements Knowledge}.
  IEEE, 1979.

\bibitem{maartensson2016continuous}
T.~M{\aa}rtensson, D.~St{\aa}hl, and J.~Bosch.
\newblock Continuous integration applied to software-intensive embedded systems
  -- problems and experiences.
\newblock In {\em International Conference on Product-Focused Software Process
  Improvement}. Springer, 2016.

\bibitem{musuvathi2008finding}
M.~Musuvathi, S.~Qadeer, T.~Ball, G.~Basler, P.~A. Nainar, and I.~Neamtiu.
\newblock Finding and reproducing heisenbugs in concurrent programs.
\newblock In {\em Symposium on Operating Systems Design and Implementation}.
  USENIX, 2008.

\bibitem{ostrand1984collecting}
T.~J. Ostrand and E.~J. Weyuker.
\newblock Collecting and categorizing software error data in an industrial
  environment.
\newblock {\em Journal of Systems and Software}, 4(4):289--300, 1984.

\bibitem{privault2013understanding}
N.~Privault.
\newblock {\em Understanding Markov chains: examples and applications}.
\newblock Springer Science \& Business Media, 2013.

\bibitem{runeson2012}
P.~Runeson, M.~Höst, A.~Rainer, and B.~Regnell.
\newblock {\em Case study research in software engineering: Guidelines and
  examples}.
\newblock John Wiley \& Sons, 2012.

\bibitem{shahin2017continuous}
M.~Shahin, M.~A. Babar, and L.~Zhu.
\newblock Continuous integration, delivery and deployment: A systematic review
  on approaches, tools, challenges and practices.
\newblock {\em IEEE Access}, 5:3909--3943, 2017.

\bibitem{strandberg-nutshell}
P.~E. Strandberg, W.~Afzal, T.~Ostrand, E.~Weyuker, and D.~Sundmark.
\newblock Automated system level regression test prioritization in a nutshell.
\newblock {\em IEEE Software}, 34(1):1--10, 2017.

\bibitem{strandberg2018trdb}
P.~E. Strandberg, W.~Afzal, and D.~Sundmark.
\newblock Decision making and visualizations based on test results.
\newblock In {\em International Symposium on Empirical Software Engineering and
  Measurement}. ACM/IEEE, 2018.

\bibitem{strandberg2019flow}
P.~E. Strandberg, E.~P. Enoiu, W.~Afzal, D.~Sundmark, and R.~Feldt.
\newblock Information flow in software testing -- an interview study with
  embedded software engineering practitioners.
\newblock {\em IEEE Access}, 7:46434--46453, 2019.

\bibitem{strandberg2018automated}
P.~E. Strandberg, T.~J. Ostrand, E.~J. Weyuker, D.~Sundmark, and W.~Afzal.
\newblock Automated test mapping and coverage for network topologies.
\newblock In {\em International Symposium on Software Testing and Analysis}.
  ACM, 2018.

\bibitem{strandberg2016}
P.~E. Strandberg, D.~Sundmark, W.~Afzal, T.~J. Ostrand, and E.~J. Weyuker.
\newblock Experience report: Automated system level regression test
  prioritization using multiple factors.
\newblock In {\em International Symposium on Software Reliability Engineering}.
  IEEE, 2016.

\bibitem{sycofyllos2016empirical}
N.~Sycofyllos.
\newblock An empirical exploration in the study of software-related fatal
  failures, 2016.
\newblock Bachelor thesis, M{\"a}lardalen University.

\bibitem{thorve2018empirical}
S.~Thorve, C.~Sreshtha, and N.~Meng.
\newblock An empirical study of flaky tests in android apps.
\newblock In {\em International Conference on Software Maintenance and
  Evolution}. IEEE, 2018.

\bibitem{vahabzadeh2015empirical}
A.~Vahabzadeh, A.~M. Fard, and A.~Mesbah.
\newblock An empirical study of bugs in test code.
\newblock In {\em International Conference on Software Maintenance and
  Evolution}. IEEE, 2015.

\bibitem{vandeursen2001refactoring}
A.~van Deursen, L.~Moonen, A.~van Den~Bergh, and G.~Kok.
\newblock Refactoring test code.
\newblock In {\em International conference on extreme programming and flexible
  processes in software engineering}, 2001.

\bibitem{wiklund2017impediments}
K.~Wiklund, S.~Eldh, D.~Sundmark, and K.~Lundqvist.
\newblock Impediments for software test automation: A systematic literature
  review.
\newblock {\em Software Testing, Verification and Reliability}, 2017.

\bibitem{wolf1994codesign}
W.~H. Wolf.
\newblock Hardware-software co-design of embedded systems.
\newblock {\em Proceedings of the IEEE}, 82(7):967--989, 1994.

\bibitem{zhang2014empirically}
S.~Zhang, D.~Jalali, J.~Wuttke, K.~Mu{\c{s}}lu, W.~Lam, M.~D. Ernst, and
  D.~Notkin.
\newblock Empirically revisiting the test independence assumption.
\newblock In {\em International Symposium on Software Testing and Analysis}.
  ACM, 2014.

\end{thebibliography}
\end{small}

\end{document}